PAPER • OPEN ACCESS

# Comprehensive topography characterization of polycrystalline diamond coatings

View the article online for updates and enhancements.

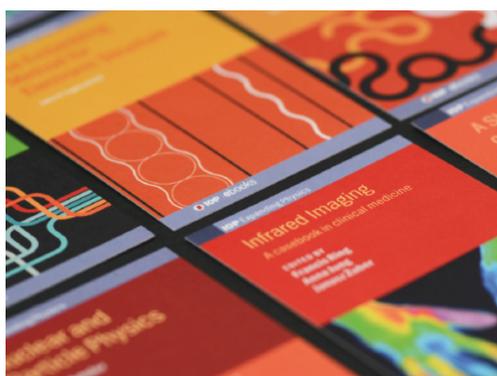

# Surface Topography: Metrology and Properties

PAPER

# Comprehensive topography characterization of polycrystalline diamond coatings

Abhijeet Gujrati[1] , Antoine Sanner[2,3], Subarna R. Khanal[1], Nicolaie Moldovan[4] , Hongjun Zeng[4,5] , Lars Pastewka[2,3,*] and Tevis D. B. Jacobs[1,*]

[1]  Department of Mechanical Engineering and Materials Science, University of Pittsburgh, 3700 O'Hara St, Pittsburgh, PA, 15261, United States of America
[2]  Department of Microsystems Engineering, University of Freiburg, Georges-Köhler-Allee 103, 79110 Freiburg, Germany
[3]  Cluster of Excellence livMatS, Freiburg Center for Interactive Materials and Bioinspired Technologies, University of Freiburg, Georges-Köhler-Allee 105, 79110 Freiburg, Germany
[4]  Alcorix Co., 14047 Franklin Ct., Plainfield, IL 60544, United States of America
[5]  Aqua via Rock LLC, 712 Muirhead Ct., Naperville, IL 60565, United States of America
*  Authors to whom any correspondence should be addressed.

E-mail: lars.pastewka@imtek.uni-freiburg.de and tjacobs@pitt.edu

## Abstract

The surface topography of diamond coatings strongly affects surface properties such as adhesion, friction, wear, and biocompatibility. However, the understanding of multi-scale topography, and its effect on properties, has been hindered by conventional measurement methods, which capture only a single length scale. Here, four different polycrystalline diamond coatings are characterized using transmission electron microscopy to assess the roughness down to the sub-nanometer scale. Then these measurements are combined, using the power spectral density (PSD), with conventional methods (stylus profilometry and atomic force microscopy) to characterize all scales of topography. The results demonstrate the critical importance of measuring topography across all length scales, especially because their PSDs cross over one another, such that a surface that is rougher at a larger scale may be smoother at a smaller scale and vice versa. Furthermore, these measurements reveal the connection between multi-scale topography and grain size, with characteristic scaling behavior at and slightly below the mean grain size, and self-affine fractal-like roughness at other length scales. At small (subgrain) scales, unpolished surfaces exhibit a common form of residual roughness that is self-affine in nature but difficult to detect with conventional methods. This approach of capturing topography from the atomic- to the macro-scale is termed *comprehensive topography characterization*, and all of the topography data from these surfaces has been made available for further analysis by experimentalists and theoreticians. Scientifically, this investigation has identified four characteristic regions of topography scaling in polycrystalline diamond materials.

## 1. Introduction

Surface topography controls surface properties of carbon coatings. For example, prior measurements of diamond-like carbon (DLC) coatings show that tribological behavior [1] and adhesion [2] are strongly affected by surface texture, all the way down to the nanoscale. The surface topography of diamond coatings [3] affects their performance, including their friction [4], wear [5], adhesion [6, 7], and biocompatibility [8]. Diamond is an important material that is used in many industrial applications [9] such as for mechanical seals [10], MEMS devices [11], biomedical applications [12], seals and bearings [13], and nuclear fusion [14] because it has low friction and wear [15, 16], and because it is robust and chemically inert so that it can be operated in corrosive environments.

Numerical models have been proposed to describe the effect of surface roughness on contact and adhesion [17]. Initially, the classic Greenwood-Williamson [18] and Fuller-Tabor [19] models described contact area and adhesion based on the average height of the roughness (figure 1(a)). Recently, the multi-scale nature of roughness has been included in the modeling of surface properties [20–29]. In particular, it has been shown that a critical quantity controlling contact area, adhesion, friction, and wear is the root-mean-square slope of the surface $h'_{rms}$ and that this quantity, for a surface with multi-scale roughness, is strongly influenced by small-scale features [22, 30]. Therefore, understanding and prediction of the topography-dependent surface properties of carbon coatings requires multi-scale characterization of roughness, down to the atomic scale.

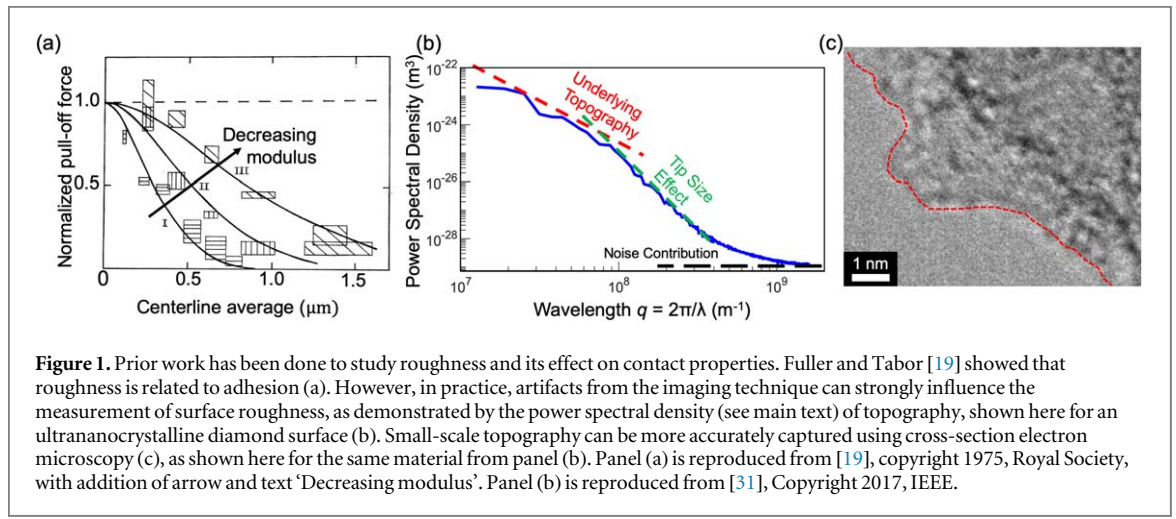

**Figure 1.** Prior work has been done to study roughness and its effect on contact properties. Fuller and Tabor [19] showed that roughness is related to adhesion (a). However, in practice, artifacts from the imaging technique can strongly influence the measurement of surface roughness, as demonstrated by the power spectral density (see main text) of topography, shown here for an ultrananocrystalline diamond surface (b). Small-scale topography can be more accurately captured using cross-section electron microscopy (c), as shown here for the same material from panel (b). Panel (a) is reproduced from [19], copyright 1975, Royal Society, with addition of arrow and text 'Decreasing modulus'. Panel (b) is reproduced from [31], Copyright 2017, IEEE.

Because conventional methods for topography measurement cannot capture these small scales, then researchers must choose either to ignore the small-scale topography, or to extrapolate it from the single-scale measurements at larger scales. Because many real-world surfaces exhibit hierarchical topography [32–36], this extrapolation is often done by assuming that the material is self-similar or self-affine. *Self-similarity* implies that the topography is statistically indistinguishable at all magnifications $\zeta$; in other words, if the lateral length scale $L$ is rescaled to $\zeta L$, then the measured height $h$ is rescaled to $\zeta h$. *Self-affinity* is related, but characterized by the Hurst exponent $H$; where rescaling the length to $\zeta L$ and the height to $\zeta^H h$ yields statistically indistinguishable surfaces. Mathematically, the root-mean-square slope $h'_{rms}$ of a self-affine, randomly-rough surface depends strongly on the smallest scales at which the roughness exists [22, 30]. For real surfaces, it is often assumed that surfaces are self-affine down to the atomic scale, but this assumption is largely untested.

Instead of ignoring or extrapolating for small-scale topography, new approaches are required for characterizing topography across all length scales. While stylus profilometry and atomic force microscopy (AFM) are indispensable tools in characterizing surface topography, these tip-based techniques are unable to provide the small-scale topography of rough surfaces because the radius of the scanning tip introduces artifacts [37]. This limits the range of reliability of the roughness measurements [30] (figure 1(b)). Optical techniques, such as scanning white-light interferometry or laser confocal microscopy, suffer from diffraction-limited lateral resolution and optical-transfer-function artifacts, and are thus similarly incapable of measuring the smallest-scale topography [38]. The result of this is that conventional measurements of surface topography are incomplete, and computed surface metrics (such as $h'_{rms}$) are unreliable, depending explicitly on the lateral resolution of the measurement. Even very advanced methods of *analyzing* surface topography, such as those in [39], are limited in their effectiveness by the range of size-scales in the underlying topography measurement. Instead, cross-section electron microscopy provides a reliable method to characterize surface topography down to the Ångström-scale [40] (figure 1(c)). Further, the small-scale topography can be stitched together with the medium- and large-scale topography using the power spectral density (PSD), to provide a comprehensive statistical description of surface topography at all size scales [32]. The PSD is a mathematical tool which separates the contribution to roughness from different length scales $\lambda$, and it is commonly represented as a function of wavevector $q = 2\pi/\lambda$.

This method of combining many different measurements, from the atomic to the macroscale, is termed *comprehensive topography characterization* and was applied here to investigate the surface roughness of four different varieties of diamond coatings, namely ultrananocrystalline diamond (UNCD), polished UNCD (pUNCD), nanocrystalline diamond (NCD), and microcrystalline diamond (MCD).

## 2. Methods

Thin films of the diamond materials were deposited (Advanced Diamond Technologies, Romeoville, IL) using a tungsten hot-filament chemical vapor deposition (HF-CVD) system with parameters as described in [41]. To improve electrical conductivity, all materials were boron doped with a B/C ratio of 3000 ppm. All materials were deposited to a thickness of 2 μm on polished silicon wafers after the wafers were sonicated with slurries containing suspended diamond nanoparticles. The deposition parameters are listed in table 1.

**Table 1.** Deposition parameters for various forms of polycrystalline diamond.

| Diamond type | CH$_4$/H$_2$ Ratio | Pressure (Torr) | Filament temperature (°C) | Filament power (KW) | Polished |
| --- | --- | --- | --- | --- | --- |
| MCD | 1.5% | 25 | 2460 | 15.4 | N |
| NCD | 2.9% | 10 | 2505 | 15.0 | N |
| UNCD | 4.7% | 5 | 2550 | 15.1 | N |
| pUNCD | 4.7% | 5 | 2550 | 15.1 | Y |

The smallest-scale topography measurements were made using transmission electron microscopy, following the techniques described in [40]. For the UNCD, NCD, and MCD, the 'wedge deposition technique' was used, whereas for pUNCD, the 'surface-preserving cross-section technique' was used. These techniques are described in detail in [40]. Briefly, the wedge deposition technique involves depositing the diamond film directly onto TEM-ready silicon thin-wedge substrates. The surface-preserving cross-section technique utilizes conventional methods for the preparation of TEM cross-sections (sectioning, grinding, polishing, dimple-grinding, and ion etching) with process modifications to ensure that the original surface is preserved. The samples were imaged using a TEM (JEOL JEM 2100F, Tokyo, Japan) operated at 200 keV. The images were taken using magnification levels from 5000× to 600 000×.

The profiles were extracted from the TEM images by using custom Matlab scripts to trace the outermost boundary of the material. Before tracing, the images were rotated such that the boundary was approximately horizontal. The vast majority of the measured surfaces were functions, i.e. for every point on the x-axis, there was exactly one point on the y-axis; in other words, the measured topographies were not reentrant. However, there were some cases where two adjacent points were captured with identical or decreasing horizontal position (locally reentrant). For these cases, the latter point was removed to restore non-reentrant behavior so that the topography can be described by a function as required by the calculation of the PSD. They were only observed in 12 out of 160 TEM profiles, and even then, only in a small number of points per profile, therefore the process of removing such points should not affect the accuracy of the analysis. Further, these locally reentrant points are attributed to imperfect rotation of the TEM profiles, rather than to truly reentrant features on the surface.

The medium-scale topography was measured using an atomic force microscope (Dimension V, Bruker, Billerica, MA) in tapping mode with diamond-like carbon-coated probes (Tap DLC300, Mikromasch, Watsonville, CA). For all substrates, square measurements were taken with the following lateral sizes: 3 scans each at 100 nm, 500 nm, and 5 μm; 1 scan each at 250 nm and 1 μm. The scanning speed was maintained at 1 μm s$^{-1}$ for all scans. Each scan had 512 lines, with 512 data points per line, corresponding to pixel sizes in the range of 0.2 to 98 nm. The wear of the AFM tip was minimized using the best practices described in [42]. Specifically, the values of free-air amplitude and amplitude ratio, which is the ratio of the amplitude of AFM probe tip vibration when performing a scan to the amplitude when vibrating in free air, were kept in the range of 37–49 nm and 0.15–0.3, respectively. Though AFM provides a two-dimensional description of the surface topography, the data were analyzed as a series of line scans. This practice maintained consistency with the other techniques, which yield one-dimensional measurements, and also eliminated artifacts due to instrumental drift in the slow-scan axis.

The largest scales of topography were measured using a stylus profilometer (Alpha Step IQ, KLA Tencor, Milpitas, CA) with a 5-μm diamond tip. Measurements were collected at a scanning speed of 10 μm s$^{-1}$, with data points every 100 nm. A total of 8 measurements were taken on each substrate, with 2 measurements each at scan sizes of 0.5, 1, 2, and 5 mm. All measurements were corrected using a parabolic fit to remove the tilt of the sample and the bowing artifact from the tool. For the UNCD and pUNCD, the larger scan sizes exhibited consistent non-parabolic trends due to instrument artifacts. These artifacts were corrected by taking reference scans on polished silicon wafers and subtracting the averaged reference profiles from the measurements.

Finally, the PSD was used to combine all measurements from a single surface into one averaged curve that describes the topography of that surface. The PSD is the Fourier transform of the autocorrelation function of a line scan with height $h(x)$, which is mathematically equivalent to the square of the amplitude of $\tilde{h}(q)$; i.e., $C(q) = L^{-1} |\tilde{h}(q)|^2$, $q$ is the wavevector and $L$ is the length of the scan. All data were collected and analyzed as 1D line scans, enabling the calculation of the one-dimensional PSD, denoted here as $C$ (designated $C^{1D}$ in [30, 32]). These calculations follow the standards established in [30] for computing and reporting PSDs.

## 3. Results

All topography data sets collected from this investigation are freely available for download and analysis [43–46].

### 3.1. Multi-scale topography measurement

Representative images of all three techniques are shown for the four materials in figure 2. The stylus profilometry data are shown with decreasing scan size in figure 2(a). It is clear that the roughness on the MCD has the largest amplitude (RMS height) while the pUNCD surface shows the smallest. Going from larger scans (figure 2(a$_1$)) to smaller scans (figure 2 (a$_3$)), the amplitude of measured topography decreases for all four diamond species. For example, while a 5-mm scan of MCD spans a vertical range of 646 nm, a 0.5-mm scan of the same surface spans just 391 nm. In order to interpret the stylus data correctly, an estimate of the tip radius is needed because the tip introduces artifacts at and below this size scale. Figure 2(a$_4$) shows a scanning electron microscopy (SEM) image of the stylus tip. Fitting a circle to the tip yields a radius of $R = 5.1$ μm. The exact point where tip artifacts become dominant [37] will be estimated using the PSD in the following section.

The AFM measurements are shown as topography maps of increasing resolution in figures 2(b)–(e). The NCD and MCD clearly show grain structure and faceting at scan sizes of 5 μm and 1 μm (figure 2(d$_{1-2}$) and (e$_{1-2}$)). The UNCD sample (figure 2(c$_{1-2}$)) also shows strong texture, but the size indicates that these are multi-grain clusters. The pUNCD shows topographic features, but no grain structure. At the highest resolution (figures 2(b$_3$)–(e$_3$)), all features look relatively smooth. The smoothness of the features is likely related to tip artifacts (caused by convolution between tip curvature and topography). Figures 2(b$_4$)–(e$_4$) show TEM images of the AFM tips after they were used to image the surfaces. The tip radii were measured as $R = 17$ nm for pUNCD, $R = 36$, 47 and 31 nm for UNCD, NCD and MCD respectively using the same procedure used for the stylus tip. As with the stylus tip, a circle was fitted to the tip profile for the extraction of the radius (figures 2(b$_4$), (c$_4$), (d$_4$), and (e$_4$)). In cases where the tip apex did not appear perfectly circular, a best-fit circle was fitted to the region of the tip that makes contact with the substrate.

Finally, the surfaces were analyzed using side-view TEM (figures 2(f)–(i)). Once again, the pUNCD surfaces had the lowest amplitude of topography and the MCD had the highest. The NCD and MCD materials showed clear faceting from the individual crystallites. At the scales accessible by the TEM, the UNCD surface also shows faceting (see figures 2(g$_2$), and (g$_3$)). However, the smooth pUNCD surface shows no indication of faceting but rather a smoothly varying surface topography. Surprisingly, despite significant differences in topography at larger scales, the smallest-scale topography is nearly identical between the microcrystalline diamond, the nanocrystalline diamond, and the ultra-nanocrystalline diamond. This finding is further discussed in the next paragraph.

To further investigate the similarities in roughness at the smallest scales, representative images of the three unpolished surfaces are shown in greater detail in figure 3. In all cases, significant roughness is visible on the scale of Angstroms to nanometers. This small-scale topography is visible even on a single facet of a single grain of the NCD and MCD materials. The atomic lattice of the diamond is clearly visible in the TEM whenever a grain is aligned with a zone axis lying near to the imaging axis. In these cases, the lattice is observed in many areas to extend to within 1 nm of the surface. Therefore, it is not simply a rough and potentially amorphous surface layer that is sitting on a flat diamond facet, rather the diamond crystal itself exhibits significant roughness at the small scale.

### 3.2. Computing topography metrics, in real-space and in frequency-space

Now, a more quantitative analysis of the topography data is presented, and scalar roughness parameters are computed for the various surfaces. First, the root-mean-square (RMS) height $h_{\text{rms}}$, RMS slope $h'_{\text{rms}}$, and RMS curvature $h''_{\text{rms}}$ are computed in real-space from each line scan by numerically integrating the squared height data (or its derivatives) over the scan length $L$ [32]:

$$h_{\text{rms}}^2 = \frac{1}{L}\int_0^L h^2(x)\,dx,\ h'^2_{\text{rms}} = \frac{1}{L}\int_0^L \left(\frac{dh}{dx}\right)^2 dx,$$
$$h''^2_{\text{rms}} = \frac{1}{L}\int_0^L \left(\frac{d^2h}{dx^2}\right)^2 dx \quad (1)$$

using the trapezoidal rule (equation (4) of [32]). Figure 4 shows the computed roughness parameters as a function of scan size $L$ (for RMS height, which depends on the larger-scale features) and measurement resolution $l$ (for RMS slope and curvature, which depend on smaller-scale features). Note that for TEM data, the 'measurement resolution' is different from the size of a pixel in the camera ('pixel size') (which can be sub-atomic at the highest magnifications). The measurement resolution $l$ is determined from the point spacing of the extracted profiles, as shown in figure 2. Computing the RMS height as a function of size is equivalent to an analysis of the surfaces' self-affine properties using a variable bandwidth method [47].

Figure 4(a) shows that $h_{\text{rms}}$ increases with $L$ for all the surfaces studied here at small $L$, i.e. $L$ less than 1 μm. There is a crossover to constant (independent of $L$) $h_{\text{rms}}$ at a scan size of 1–10 μm. All of the surfaces studied here show this transition, which corresponds to the thickness of these coatings (2 μm). The amplitude of the pUNCD surface is much smaller than the other three surfaces. NCD and MCD roll off to the same constant value. The values in figure 4(a) at large $L$ correspond to the observation of the amplitude in stylus profilometry shown in figure 2, with pUNCD being the 'smoothest' and MCD being the 'roughest' surface at large scales.

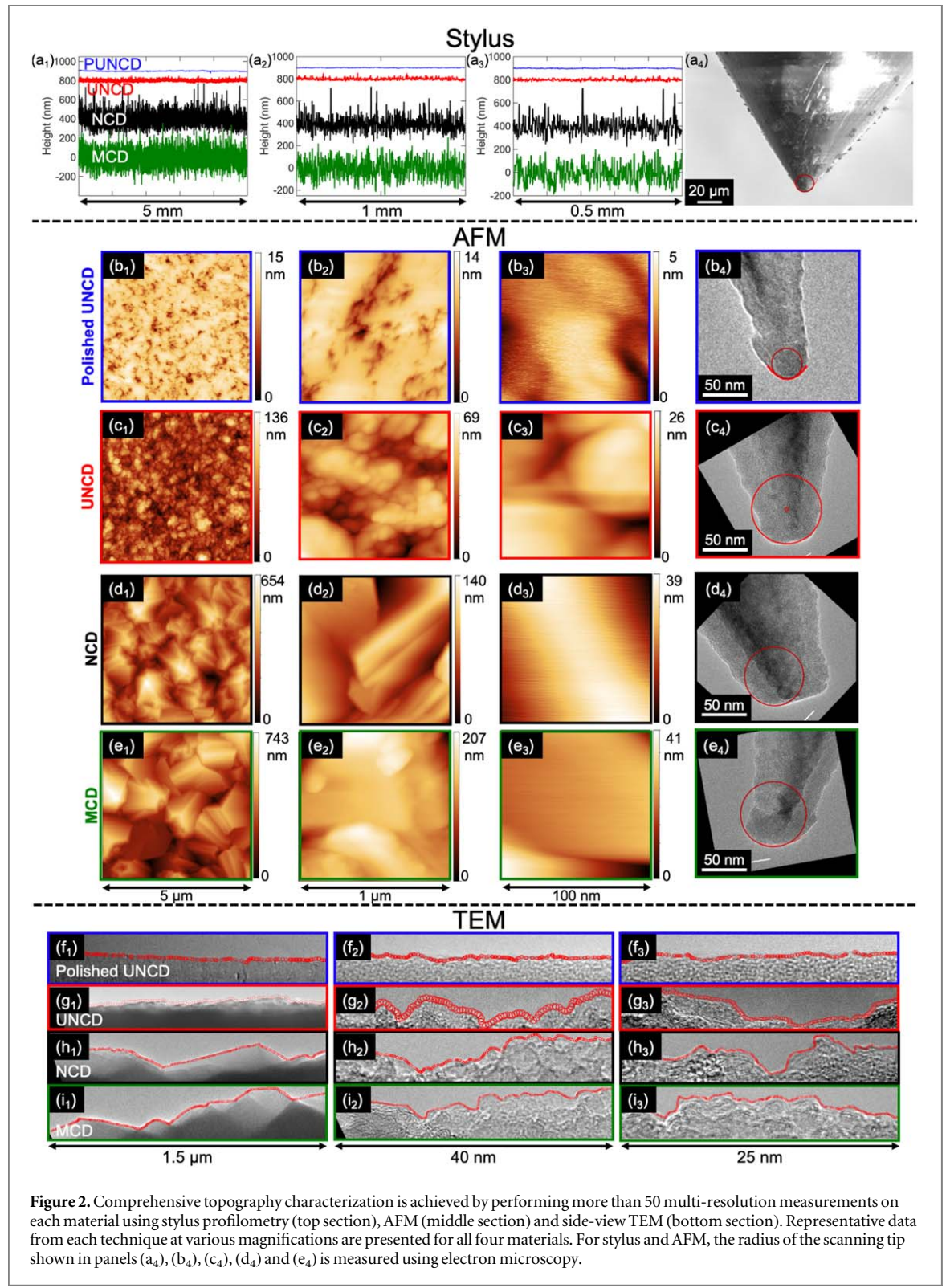

**Figure 2.** Comprehensive topography characterization is achieved by performing more than 50 multi-resolution measurements on each material using stylus profilometry (top section), AFM (middle section) and side-view TEM (bottom section). Representative data from each technique at various magnifications are presented for all four materials. For stylus and AFM, the radius of the scanning tip shown in panels (a$_4$), (b$_4$), (c$_4$), (d$_4$) and (e$_4$) is measured using electron microscopy.

From the analysis of ideal self-affine surfaces [22, 30, 32], it can be shown that $h_{\rm rms} \propto L^H$ while $h'_{\rm rms} \propto l^{H-1}$ and $h''_{\rm rms} \propto l^{H-2}$. Therefore, in figure 4(a), the small-scale data has been fit with a power-law function (solid lines), and the extracted values are used to determine the expected trends in figures 4(b), (c) (dashed lines). As argued above and first described in [36, 48], it is apparent from figure 4 that no single value of RMS height, RMS slope, and RMS curvature can be defined, because these parameters depend on $L$ or $l$. This demonstrates a key difficulty that impedes efforts to link surface function to a single scalar roughness parameter; these common roughness parameters (including RMS roughness) are scanning-length dependent and do not describe an intrinsic property of the material.

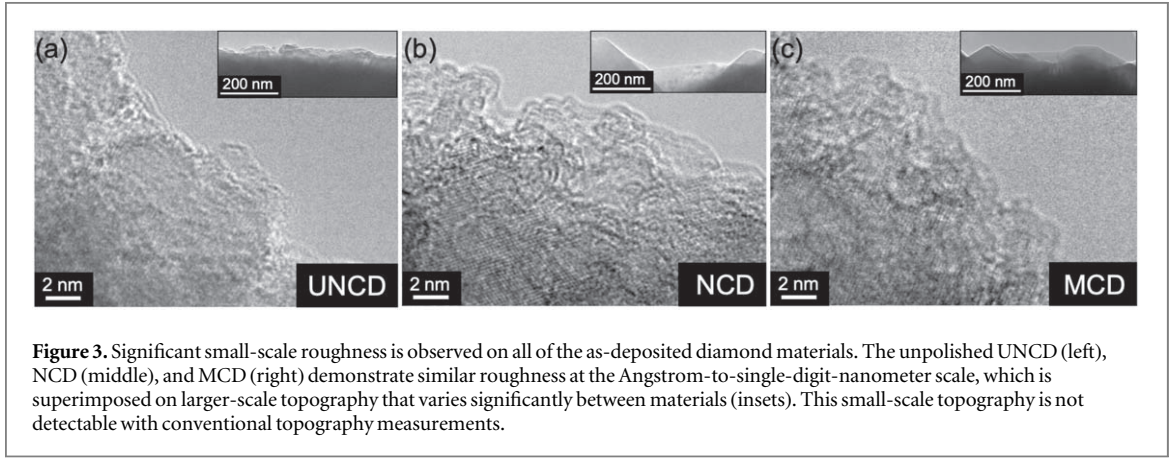

**Figure 3.** Significant small-scale roughness is observed on all of the as-deposited diamond materials. The unpolished UNCD (left), NCD (middle), and MCD (right) demonstrate similar roughness at the Angstrom-to-single-digit-nanometer scale, which is superimposed on larger-scale topography that varies significantly between materials (insets). This small-scale topography is not detectable with conventional topography measurements.

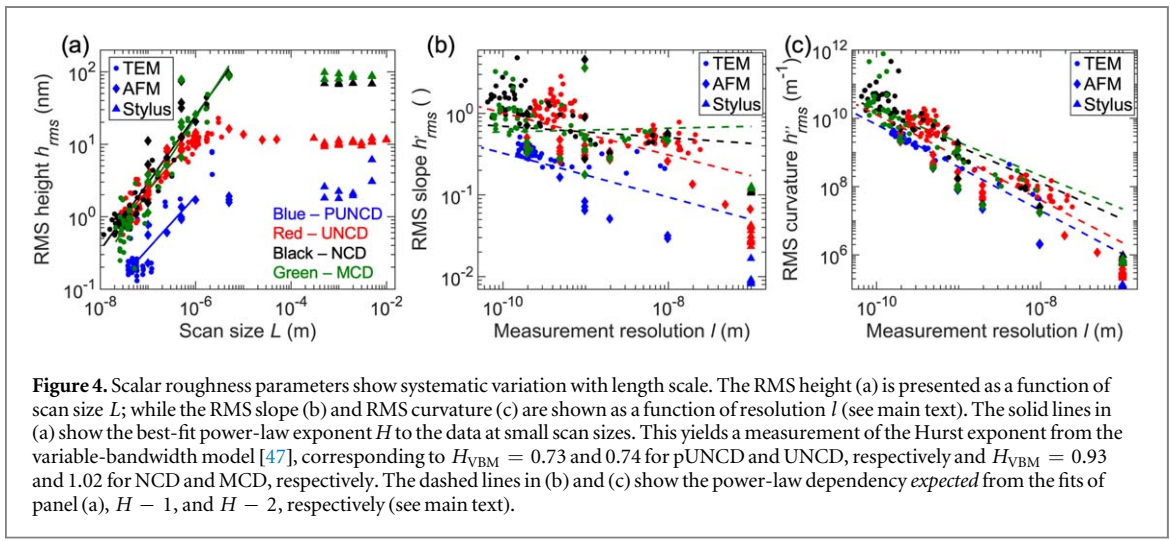

**Figure 4.** Scalar roughness parameters show systematic variation with length scale. The RMS height (a) is presented as a function of scan size $L$; while the RMS slope (b) and RMS curvature (c) are shown as a function of resolution $l$ (see main text). The solid lines in (a) show the best-fit power-law exponent $H$ to the data at small scan sizes. This yields a measurement of the Hurst exponent from the variable-bandwidth model [47], corresponding to $H_{VBM} = 0.73$ and 0.74 for pUNCD and UNCD, respectively and $H_{VBM} = 0.93$ and 1.02 for NCD and MCD, respectively. The dashed lines in (b) and (c) show the power-law dependency *expected* from the fits of panel (a), $H-1$, and $H-2$, respectively (see main text).

As an alternative to analyzing the real-space measurements individually (as done in figure 4), the measurements can all be combined in frequency space to create a single PSD, denoted $C(q)$, that yields a complete statistical description of topography for each surface, as shown in figure 5. This analysis was carried as follows: first, the PSD of each measurement was computed, following the procedures laid out in [30, 32]. Second, a reliability cutoff [30, 32] due to tip artifacts was calculated for each PSD based on the measured tip radius (figure 2), and all data below this size scale was deemed unreliable and removed [32]. Third, all of the reliable portions of the many individual PSDs were combined by computing the arithmetic average of all measurements in logarithmically-spaced bins. The result is a single whole-surface PSD that describes the material across all length scales. There are no fitting parameters in this analysis; rather the PSD serves to separate the different size scales of topography, and the various techniques agree within experimental uncertainty. The only exception is for pUNCD, where the small-size stylus data lies below the AFM and TEM data, causing a dip around $q = 10^7$ m$^{-1}$. This is believed to be an instrumental artifact, rather than resulting from the real topography. Overall, the value of these comprehensive PSDs is that they can be used in analytical and numerical models (such as [22, 25–27, 49–52]) to understand and predict surface properties (e.g. [17]).

To compute scale-invariant scalar roughness parameters, the full stitched-together PSD was used to compute RMS height, RMS slope, and RMS curvature as:

$$(h_{rms})^2 = \frac{1}{\pi}\int_0^\infty C(q)\,dq, \quad (h'_{rms})^2 = \frac{1}{\pi}\int_0^\infty q^2 C(q)\,dq,$$
$$(h''_{rms})^2 = \frac{1}{\pi}\int_0^\infty q^4 C(q)\,dq$$

(2)

Table 2 shows the computed RMS parameters. Also computed from the whole-surface PSD is the area ratio, which is the increase in full surface area, per unit apparent area, due to the roughness. The area ratio is a critical parameter in investigations of soft-material adhesion, including [17], and it is calculated as described in [17]. While MCD is the roughest in terms of RMS height, the unpolished UNCD is the steepest in terms of RMS slope. However, while table 2 shows the mathematically correct or 'true' values of RMS parameters for a surface, any individual application or surface property may depend only on a certain range of length-scales. In that case, scale-dependent parameters would be recomputed by integrating only across the relevant size scales.

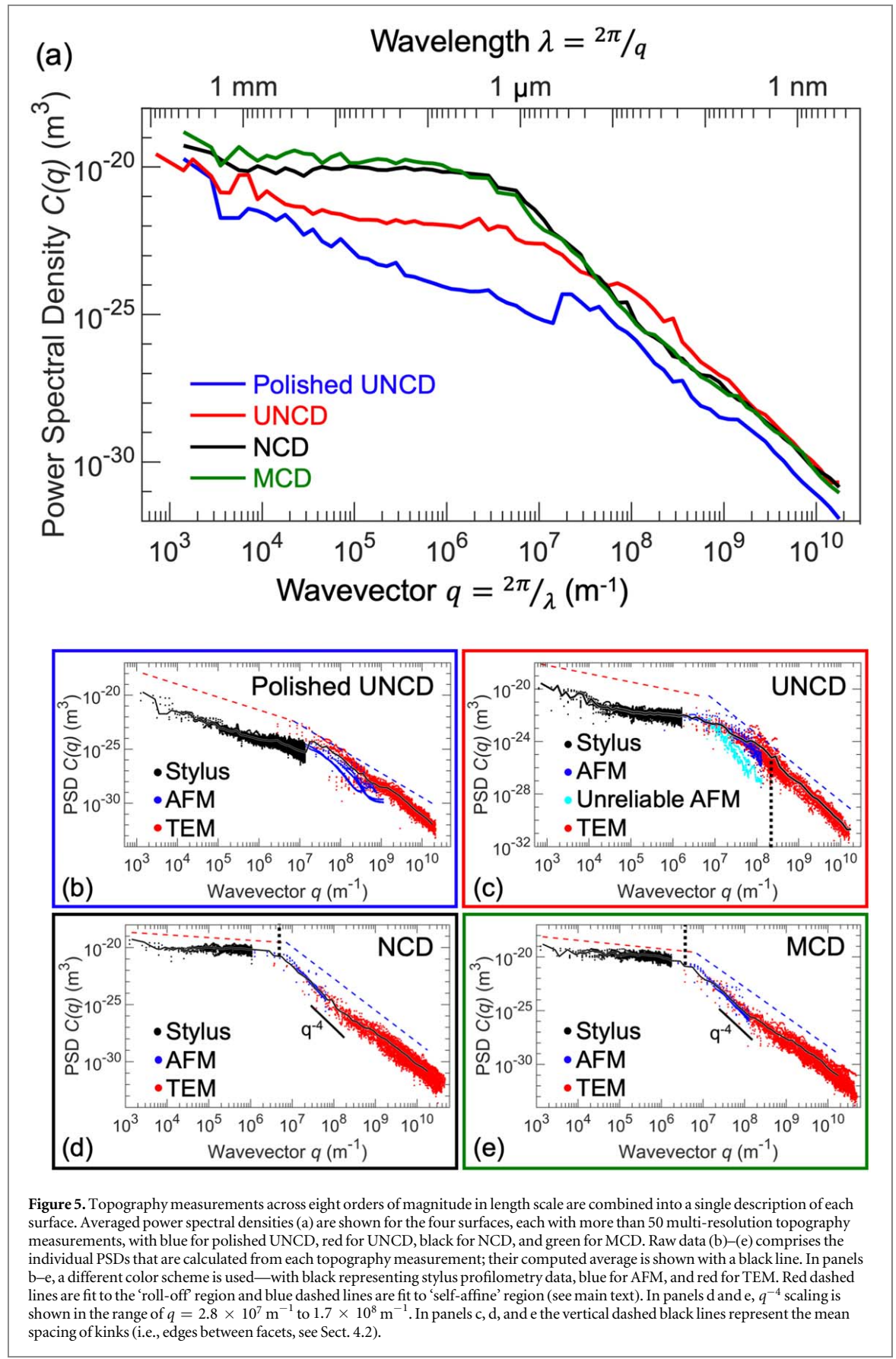

**Figure 5.** Topography measurements across eight orders of magnitude in length scale are combined into a single description of each surface. Averaged power spectral densities (a) are shown for the four surfaces, each with more than 50 multi-resolution topography measurements, with blue for polished UNCD, red for UNCD, black for NCD, and green for MCD. Raw data (b)–(e) comprises the individual PSDs that are calculated from each topography measurement; their computed average is shown with a black line. In panels b–e, a different color scheme is used—with black representing stylus profilometry data, blue for AFM, and red for TEM. Red dashed lines are fit to the 'roll-off' region and blue dashed lines are fit to 'self-affine' region (see main text). In panels d and e, $q^{-4}$ scaling is shown in the range of $q = 2.8 \times 10^7 \, \text{m}^{-1}$ to $1.7 \times 10^8 \, \text{m}^{-1}$. In panels c, d, and e the vertical dashed black lines represent the mean spacing of kinks (i.e., edges between facets, see Sect. 4.2).

**Table 2.** One-dimensional roughness parameters for nanodiamond substrates computed from the whole-surface PSD for these materials.

|  | Polished UNCD | UNCD | NCD | MCD |
| --- | --- | --- | --- | --- |
| RMS height | $4.2 \pm 0.8$ nm | $17.4 \pm 1.3$ nm | $97.2 \pm 11.7$ nm | $101.2 \pm 8.0$ nm |
| RMS slope | $0.31 \pm 0.03$ | $1.17 \pm 0.28$ | $0.92 \pm 0.10$ | $0.85 \pm 0.10$ |
| RMS curvature | $1.99 \pm 0.35$ nm$^{-1}$ | $6.32 \pm 1.20$ nm$^{-1}$ | $5.91 \pm 1.83$ nm$^{-1}$ | $5.04 \pm 1.45$ nm$^{-1}$ |
| Area Ratio | 1.07 | 1.69 | 1.47 | 1.42 |

## 4. Discussion

### 4.1. Evaluating the fractal nature of diamond coatings, and the meaning of Hurst exponents

The Hurst exponent (which can be related to the fractal dimension, as described in [53]), can be calculated from the PSD, which is commonly separated (somewhat arbitrarily) into the 'self-affine' region, where the topography appears to be described by a power law relationship of $C \propto q^\beta$ where $\beta$ is the power-law exponent, and the 'roll-off' region, where the PSD appears to be flatter. The Hurst exponent $H$ is typically extracted from the self-affine region as $H = (\beta - 1)/2$ [53, 54]. Using this procedure, and the region of the curves between $q = 6.3 \times 10^6$ m$^{-1}$ and $1.8 \times 10^{10}$ m$^{-1}$, the resulting value was $H = 0.62 \pm 0.09$ and $0.77 \pm 0.06$ for pUNCD and UNCD respectively; and $H = 0.89 \pm 0.04$ and $0.87 \pm 0.03$ for NCD and MCD respectively.

There are two alternative methods of extracting the Hurst exponent: from the real-space data using the variable bandwidth method (VBM) [55]; and from the roll-off region by assuming that the full PSD is described by Fractional Gaussian Noise (FGN) [53], a hypothesis put forth by some of us in [32]. The VBM is nothing more than an analysis of the functional dependence of RMS height $h_{\text{rms}}$ as a function of scan size $L$ (that scales as $h_{\text{rms}}(L) \propto L^H$ for self-affine surfaces), as shown in figure 4(a). Note that even for an individual scan, the RMS height could be computed over a subsection of that scan (yielding an estimate of the Hurst exponent for a single realization of the topography [33, 47, 55]) but here full-size scans were used for calculation. As shown in figure 4(a), RMS height $h_{\text{rms}}(L)$ can be accurately fit with a power-law form over the range from the smallest size ($L \sim 10$ nm) up to approximately $L = 1 - 10$ μm. Over this region, the Hurst exponents were $H_{\text{VBM}} = 0.73 \pm 0.18$ for pUNCD; $H_{\text{VBM}} = 0.74 \pm 0.05$ for UNCD; $H_{\text{VBM}} = 0.93 \pm 0.09$ for NCD and $H_{\text{VBM}} = 1.02 \pm 0.10$ for MCD. The Hurst exponents from this method ($H_{\text{VBM}}$) yield similar results to the above Hurst exponents ($H$) that are extracted from the more common method of fitting the 'self-affine' region of the PSD. The VBM is a useful technique because it is simpler to perform—requiring only a straightforward calculation of the root-mean-square height, calculated in real-space, from a series of multi-resolution measurements. Additionally, it can be readily performed on reentrant surfaces, while the calculation of the PSD requires surfaces to be functions (one height value for each lateral position).

The second alternative method for computing the Hurst exponent uses the 'roll-off' region and the fractional Gaussian noise (FGN) approach. Here, the Hurst exponent is given by $H_{\text{FGN}} = (\alpha + 1)/2$ [32, 53], where $\alpha$ is the scaling exponent $C(q) \propto q^\alpha$ at low $q$. This results in $H_{\text{FGN}} = 1.10 \pm 0.04$ for pUNCD; $H_{\text{FGN}} = 0.82 \pm 0.04$ for UNCD; and $H_{\text{FGN}} = 0.62 \pm 0.04$ and $0.70 \pm 0.05$ for NCD and MCD, respectively. A prior paper by the present authors ([32]) speculated that there may be a connection between $H_{\text{FGN}}$ and $H$ (from the self-affine region). This observation would be extremely useful as it suggests that the small-scale behavior could be predicted from large-scale measurements. Unfortunately, when these four different surfaces are compared, there is no clear relationship that emerges.

Even in the traditional 'self-affine' portion of the PSD, a range of values can be extracted for Hurst exponent for a single surface depending on the window used for fitting $H$. This is particularly true for MCD and NCD where Hurst exponents can be calculated by dividing the 'self-affine' region into two parts. For $q$ in the range of $2.8 \times 10^7$ to $1.7 \times 10^8$ m$^{-1}$, the extracted values are $H_{\text{larger}\lambda} = 1.27 \pm 0.27$ and $1.32 \pm 0.09$ for NCD and MCD, respectively; for q in the range of $1.7 \times 10^8$ to $1.8 \times 10^{10}$ m$^{-1}$, the extracted Hurst exponents are $H_{\text{smaller}\lambda} = 0.75 \pm 0.04$ and $0.78 \pm 0.05$ for NCD and MCD, respectively. Indeed, the roughness in these two portions of the curve seems to be qualitatively different, with the upper portion having a scaling behavior near $C \propto q^{-4}$, corresponding to $H = 1.5$. The origin of these differences in scaling behavior between different length scales is discussed in detail in the next section.

Because there can be so much variability in the measurement of a single surface, the whole practice of assuming self-affinity and assigning a single Hurst exponent to describe a surface must be done with caution. It is mathematically convenient to assume self-affinity as this simplifies numerical and analytical models, and it is common practice in experiments to use assumptions of self-affinity to extrapolate to small scales where the topography is not easily measured.

However, at least for the diamond materials investigated here, the best-fit value for *H* depends on the region over which it is measured, and it is strongly influenced by other factors such as grain size (see next section). Instead, where possible, it is preferable to measure surface topography across all size scales and to use the whole-surface PSD as the primary descriptor for the surface, rather than any scalar parameter.

### 4.2. The effect of grain size on topography

As discussed in the previous section, the larger-grain-size materials (NCD and MCD) demonstrate a region where the PSD scaling is similar to $q^{-4}$ in the larger-wavelength portion of the 'self-affine' region (see solid black lines in figures 5(d) and (e)). This scaling is characteristic of 'kinks' in the real-space line scan, such as sharp peaks or valleys. Note that such kinks can arise in topography as an artifact of the nonvanishing tip radius [30, 37], so one must first rule out this as the cause. However, in the present work, this $q^{-4}$ scaling is clearly observed both in the reliable portion of the AFM measurement and in the TEM measurement, both of which are free from tip-based artifacts. Therefore, this $q^{-4}$ scaling in the PSDs of the MCD and NCD is a feature of the measured topography, rather than emerging from an artifact. This behavior of the PSD corresponds to kinks in the surface topography that are directly observable in the TEM imaging, as shown in figures 2 and 3.

Our hypothesis to describe this local $q^{-4}$ scaling is that it is characteristic of topography at scales approximately equal to the grain size of the material. At these scales, features are dominated by the crystal facets and kinks between them: adjacent grains with different orientations will create concave kinks where the grains meet, while edges between crystal facets will create convex kinks. It is therefore assumed that typical topography line scans will pass through approximately two kinks (one concave and one convex) per grain. At sizes much below the grain size, the topography of MCD and NCD reverts to random, self-affine behavior, showing roughness very similar to UNCD (see figure 3), where the PSD scales as $q^{-1-2H}$ (see blue dashed lines in figures 5(b)–(e)). This finding agrees with prior work [56] on fracture surfaces in sandstone. In that work, stylus profilometry measurements showed a transition around the grain size; however sub-grain features were mostly inaccessible there due to the aforementioned tip artifacts. Computer-generated profiles were used to verify that facets can cause $q^{-4}$-scaling at a wavevector related to the average kink spacing, designated $l_k$. The mathematical basis for this hypothesis linking kink spacing to $q^{-4}$ scaling in the PSD is given in the appendix.

In order to demonstrate how the PSD is affected by the superposition of facets from a characteristic grain size and random roughness below that size, artificial one-dimensional surfaces were created that were composed of: a superposition of triangular peaks (figure 6($a_1$)); self-affine random roughness (figure 6($a_2$)); and the summation of those two into a single surface (figure 6($a_3$)). The piecewise linear surface (figure 6($a_1$)) has kinks with uncorrelated heights drawn from a Gaussian distribution and lateral distances between kinks drawn from a Rayleigh distribution. The surface is scaled in order to have an RMS slope of 1 and an average kink spacing $l_k$. The small-scale self-affine random roughness (figure 6($a_2$)) is generated using a Fourier-filtering algorithm [30, 57] with a Hurst exponent $H = 0.8$ and RMS-slope of 1.2.

When the PSDs (figure 6(b)) are computed from these surfaces, there is a transition from a flat PSD at the largest sizes to scaling as $C \propto q^{-4}$, and the transition point occurs at a wavelength of $\lambda_k = 4l_k$, which corresponds to $q_k = \frac{2\pi}{\lambda_k} = \frac{\pi}{2l_k}$. (The transition is somewhat gradual; the choice of this particular value for the transition point is motivated from the mathematical consideration in the appendix). Importantly, the summed-surface PSD follows the kinked-surface PSD and displays $q^{-4}$ scaling at larger scales, and then transitions to self-affine scaling at smaller scales, in this case corresponding to a Hurst exponent $H$ of 0.8. To ensure that these results were not unique to the particular way that the kinked surface was generated, this analysis was repeated using kinked surfaces with uniform and exponential distributions of kink spacings, as well as a surface with slopes alternating between $-1$ and 1. These analyses are shown in the supplemental section 1 (available online at stacks.iop.org/STMP/9/014003/mmedia), but the results are similar, differing only in the sharpness of transitions. The key finding from this analysis is that the kink spacing, which corresponds approximately to the grain size, introduces a signature in the scaling of the topography that causes deviations from the commonly assumed fractal-like self-affine scaling.

The PSD of the summed surface (figure 6(b)) reproduces the three regions that are visible in the PSDs of MCD and NCD: flat behavior at small $q$ (large sizes); scaling like $q^{-4}$ at intermediate values; and self-affine scaling ($H \sim 0.8$) at large $q$ (small sizes). However, this does not yet explain the self-affine scaling behavior that is observed above the grain size in polished and unpolished UNCD. To account for this, another synthetic surface was created that is similar to the first, but this time with spatially correlated kink heights (figure 6(c)). The spatial correlation of the kink heights is enforced using the method of [56] as follows: a self-affine random surface was generated using the Fourier-filtering algorithm; then facets were computed by interpolating linearly between random points (the kinks); finally this faceted surface was summed with the same self-affine random roughness as before (figure 6($a_2$)). While this does not substantially alter the behavior at small wavelengths, this adds in an additional self-affine region above the grain size (figure 6(d)).

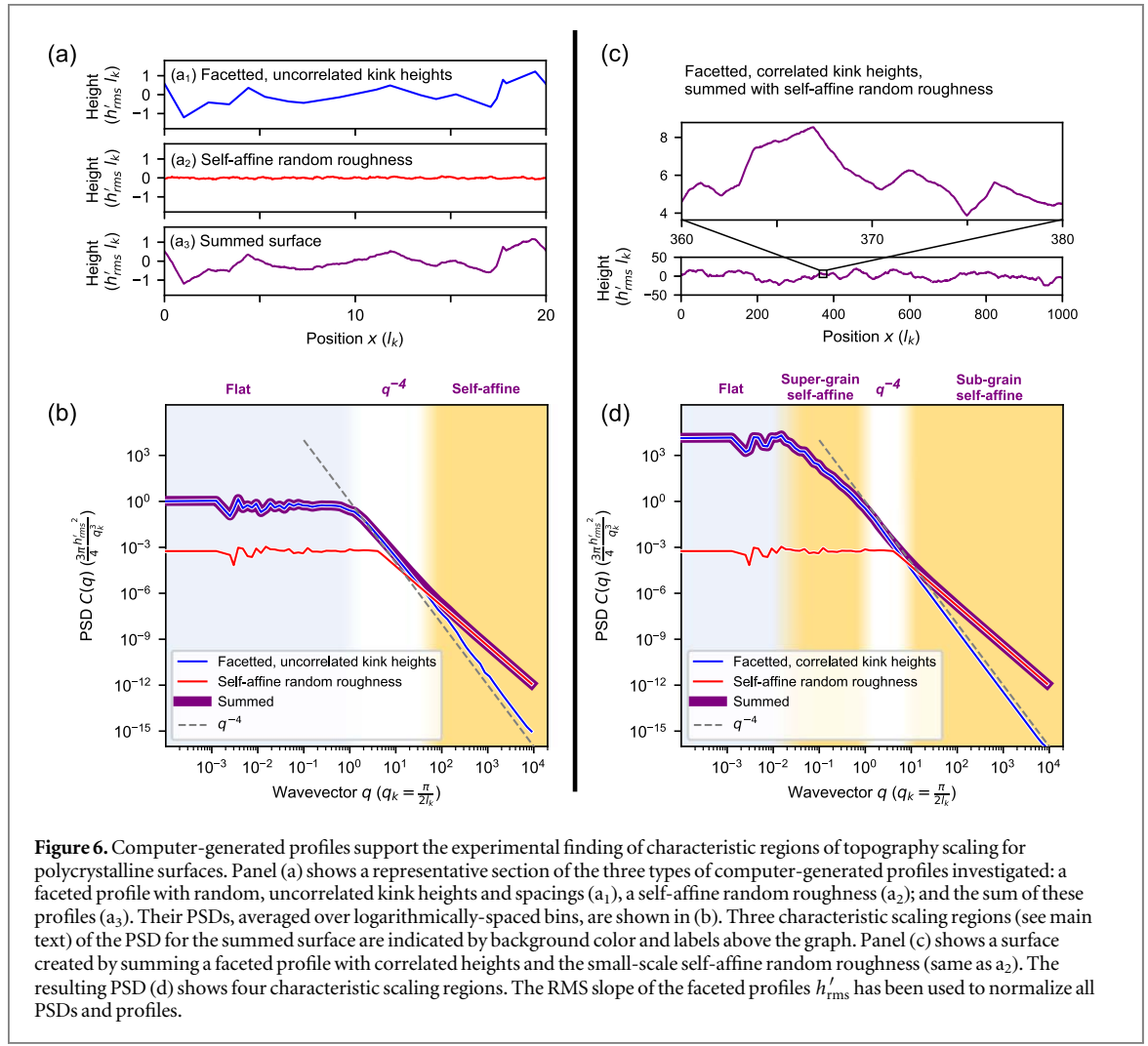

**Figure 6.** Computer-generated profiles support the experimental finding of characteristic regions of topography scaling for polycrystalline surfaces. Panel (a) shows a representative section of the three types of computer-generated profiles investigated: a faceted profile with random, uncorrelated kink heights and spacings ($a_1$), a self-affine random roughness ($a_2$); and the sum of these profiles ($a_3$). Their PSDs, averaged over logarithmically-spaced bins, are shown in (b). Three characteristic scaling regions (see main text) of the PSD for the summed surface are indicated by background color and labels above the graph. Panel (c) shows a surface created by summing a faceted profile with correlated heights and the small-scale self-affine random roughness (same as $a_2$). The resulting PSD (d) shows four characteristic scaling regions. The RMS slope of the faceted profiles $h'_{\rm rms}$ has been used to normalize all PSDs and profiles.

To verify this proposed link between $q^{-4}$ scaling of the PSD and the kink spacing of the material, an analysis of kink spacing and grain size was performed on the real materials using the AFM and TEM measurements. For the MCD and NCD, concave kinks between grains were readily visible, and therefore were quantified using feature sampling, as is used in metallographic analysis [58] (supplemental section 2.1). The mean lineal intercept between concave kinks, which is of the same order as the grain diameter [58], was measured as $839 \pm 68$ nm for MCD and $647 \pm 42$ nm for NCD. Because the mean kink spacing used in the mathematical analysis involves both convex and concave kinks, it was computed as half of this value, i.e. $419 \pm 34$ nm for MCD and $323 \pm 21$ nm for NCD. These values can be converted to frequency-space using $q_k = \frac{\pi}{2l_k}$, as discussed above, and the values calculated are $q_k = 3.7 \times 10^6$ m$^{-1}$ for MCD and $q_k = 4.9 \times 10^6$ m$^{-1}$ for NCD. The same approach could not be applied to UNCD where the kink spacing was at or below the reliability cut-off due to tip artifacts. Therefore, the mean kink spacing for UNCD was estimated as half of the grain size, which was computed by averaging a sample of 20 different grains observed in the TEM (see supplemental section 2.2). The mean grain size of the UNCD was determined to be $14 \pm 3$ nm, which corresponds to a kink spacing for UNCD of $7 \pm 2$ nm and $q_k = 2.2 \times 10^8$ m$^{-1}$. The mean kink spacings of these three materials are indicated as vertical bars in figures 5(c)–(e). Indeed, the MCD and NCD demonstrate that the scaling transitions to $C \sim q^{-4}$ right around the mean kink spacing. For UNCD, the mean kink spacing is too small to observe a clear $q^{-4}$ scaling regime; however, the self-affine scaling behavior is confirmed for sizes much larger than the characteristic grain size. The remainder of the paper will frame results in terms of grain size (again, approximately twice the value of kink spacing), since grain size is more widely measured and reported for polycrystalline materials.

In summary, based on the experimental measurements of these four polycrystalline diamond surfaces, and based on the computed PSDs of artificially generated surfaces, four characteristic regimes of topography scaling have been identified. (1) At the smallest size scales (if significantly smaller than the grain size) power-law scaling may be observed that is characteristic of self-affine random roughness, e.g., $C \sim q^{-1-2H}$, corresponding to H in the range of 0.6–0.9. (2) At size scales similar to and slightly smaller than the average grain size, the PSD displays characteristic scaling of $C \sim q^{-4}$ due to grain facets and kinks. (3) At sizes larger than the grain size, but smaller than the film thickness, there is another region of power-law scaling corresponding to random roughness ($C \sim q^{-1-2H}$). (4) Finally, at sizes larger than the film thickness, the PSD flattens out, with scaling in the range of $q^0$ to $q^{-1}$. These four regions of topography scaling accurately describe the polycrystalline diamond surfaces and computer-generated surfaces investigated here. These four regions and their boundaries may be broadly generalizable to other materials, but further investigation is required. For instance, the boundary between regions (3) and (4) corresponded to the thickness of the diamond coatings in these measurements, but this thickness was not varied to explicitly investigate this connection. The future application of comprehensive topography characterization on other materials will elucidate the applicability of these four regions to other polycrystalline materials.

## 5. Conclusions

First, these results further establish a multi-resolution approach that is designated 'comprehensive topography characterization,' which combines multiple different techniques at multiple magnifications for the same surface. Then the power spectral density can be used to combine all measurements into one statistical description of the surface. Because typical roughness metrics are inherently scale-dependent and incomplete, this paper provides a method to understand roughness at all scales, including the specific scale over which it may be relevant in a given device. Particularly the measurement of small-scale roughness may be extremely important to predict and tailor surface properties such as adhesion, friction and wear. Furthermore, all topography measurements from this publication have been made publicly available, [43–46] so that other experimentalists may compare results and so that computational modelers may use it in models to predict properties of diamond materials. The purpose of this data and the underlying approach is to advance the field towards the goal of fundamental, predictive understanding of the performance of rough surfaces.

Second, these results show that the surface roughness of polycrystalline diamond materials varies significantly with scale, with surfaces that are smoother when measured at the large-scale showing roughness that is identical or even higher when measured at the smaller scales. Furthermore, while self-affine scaling ($H \sim 0.6 - 0.9$) is observed over some lengthscales, the grain size introduces a signature into the power spectral density, showing $q^{-4}$ scaling behavior at scales approximately equal to the grain size. All unpolished surfaces show identical self-affine scaling at the smallest scale. This is a signature of small-scale roughness that is superimposed on the crystalline facets and is not observable with conventional (AFM, stylus) measurement techniques. Altogether, four characteristic regions of topography scaling were observed; these are expected to be applicable to all unpolished polycrystalline diamond films and may apply more broadly to other polycrystalline materials.

## Acknowledgments

Sample preparation was performed in the Fischione Instruments Electron Microscopy Sample Preparation Laboratory at the University of Pittsburgh. The authors acknowledge the use of two facilities at the University of Pittsburgh: the Materials Micro-Characterization Laboratory (MMCL) in the Department of Mechanical Engineering and Materials Science; and the Nanoscale Fabrication and Characterization Facility (NFCF) in the Petersen Institute of Nano Science and Engineering (PINSE). The authors thank A. Baker for supplemental TEM imaging. LP acknowledges funding by the Deutsche Forschungsgemeinschaft (DFG, German Research Foundation) under Germany's Excellence Strategy (project EXC-2193/1–390951807) and by the European Commission (ERC-StG-757343). Funding for TDBJ and the University of Pittsburgh work was provided by the National Science Foundation under award number CMMI-1727378.

## Data availability statement

The data that support the findings of this study are openly available at the following URLs: https://contact.engineering/go/wcqj3/; https://contact.engineering/go/8sc7t/; https://contact.engineering/go/cjy6s/; https://contact.engineering/go/mz7z5/.

## Appendix. A mathematical basis for the $q^{-4}$ scaling of the PSD slightly below the grain size

The analysis begins with a triangular peak $y_p(x)$:

$$y_p(x) = \begin{cases} h\dfrac{x - x_l}{x_c - x_l}, & x_l \leqslant x < x_c \\ h\left(1 - \dfrac{x - x_c}{x_r - x_c}\right), & x_c \leqslant x \leqslant x_r \\ 0, & else \end{cases} \quad (A1)$$

The Fourier transformation of the peak profile $y_p(x)$ is:

$$\tilde{y}_p(q) = \frac{h}{q^2} e^{-iqx_c} \left( \frac{1 - e^{-iq(x_r - x_c)}}{x_r - x_c} \right.$$
$$\left. + \frac{1 - e^{iq(x_c - x_l)}}{x_c - x_l} \right) \quad (A2)$$

For large $q$, the PSD of the peak $C(q) \propto |\tilde{y}(q)|^2$ oscillates with an amplitude decaying as $q^{-4}$. For $q \ll \min\left(\frac{\pi}{2(x_r - x_c)}, \frac{\pi}{2(x_c - x_l)}\right)$, $\tilde{y}_p(q) \simeq \frac{h}{2}(x_r - x_l) e^{-iqx_c}$, so the PSD is flat.

A piecewise linear function $y(x)$ with kinks at $(x_k, h_k)$ can be written as a superposition of these peaks. Its Fourier transform is:

$$\tilde{y}(q) = \sum_{k=0}^{n-1} \frac{h_k}{q^2} e^{-iqx_k} \left( \frac{1 - e^{-iq(x_{k+1} - x_k)}}{x_{k+1} - x_k} \right.$$
$$\left. + \frac{1 - e^{iq(x_k - x_{k-1})}}{x_k - x_{k-1}} \right) \quad (A3)$$

$h_n = h_0$ ensures continuity at the periodic boundary.

The PSD of the piecewise linear function has the same features as the PSD of the triangular peak: for large enough values of $q$, $C(q) \propto q^{-4}$; for $q \ll \min_k \frac{\pi}{2(x_{k+1} - x_k)}$, $\tilde{y}(q) \simeq \sum_{k=0}^{n-1} \frac{h_k(x_{k+1} - x_k)}{2} e^{-iqx_k}$ and the PSD is flat. Figure 6(b) shows that for the profiles considered, the PSD changes between flat and $\propto q^{-4}$ around $q_k = \frac{\pi}{2l_k}$, with $l_k$ the mean kink spacing, rather than the maximum kink spacing. Similarly, the PSD in figure 6(d) changes between the super-grain self-affine scaling and $\propto q^{-4}$ around the same wavevector $q_k = \frac{\pi}{2l_k}$. This is close to the value $q_k \simeq \frac{\pi}{3l_k}$ determined in [56] for the same Hurst exponent of the super-grain self-affine regime, $H = 0.8$.

The different exponents of the PSD of faceted versus self-affine surfaces correspond to different behaviors of the scale-dependent RMS slope with increasing resolution. The scale-dependent RMS slope can be computed from the PSD $C(q) = q^\alpha$ using equation (2): $h'_{\text{rms}}(q_{\max}) = \left( \int_{q_{\min}}^{q_{\max}} dq\, q^2 q^\alpha \right)^{\frac{1}{2}}$. As $q_{\max} \to \infty$, $h'_{\text{rms}}(q_{\max})$ converges to a finite value for $\alpha < -3$, but diverges for $\alpha \geqslant -3$, the former case corresponding to a faceted surface and the latter to a self-affine random surface with $H \leqslant 1$. Since the slope between two kinks is constant by definition, the RMS slope of a faceted profile is finite and reaches its limit value once the smallest resolved length is below the kink spacing. In contrast to that, the RMS slope of an ideal self-affine random surface with $H \leqslant 1$ increases indefinitely with resolution, as smaller-scale features with increasing slope are resolved.

## ORCID iDs

Abhijeet Gujrati  https://orcid.org/0000-0001-7744-5743
Nicolaie Moldovan  https://orcid.org/0000-0001-5715-4957
Hongjun Zeng  https://orcid.org/0000-0002-4415-1367
Lars Pastewka  https://orcid.org/0000-0001-8351-7336
Tevis D. B. Jacobs  https://orcid.org/0000-0001-8576-914X